# All-Digital Wideband Space-Frequency Beamforming for the SKA Aperture Array

Vasily A. Khlebnikov, 44-01865-273302, w.khlebnikov@ieee.org,
Kristian Zarb-Adami, 44-01865-273302, kza@astro.ox.ac.uk,
Richard P. Armstrong, 44-01865-273302, richard.armstrong@physics.ox.ac.uk, and
Michael E. Jones, 44-01865-273441, mike@astro.ox.ac.uk,
Department of Astrophysics, University of Oxford,
Denys Wilkinson Building, Keble Road, Oxford, OX1 3RH, United Kingdom

**Abstract**—*In this paper, we consider the problem of optimum multi-domain real-time beamforming and high-precision beam pattern positioning in application to very large wideband array antennas, particularly to the Square Kilometre Array (SKA) aperture array antenna. We present a new structure for wideband space-frequency beamforming and beamsteering that maximizes detectability of cosmic signals over the array operational frequency range.*

## 1. INTRODUCTION

The SKA is a multi-purpose radio telescope that will operate at metre and centimetere wave-lengths with extremely high sensitivity and wide field of view (FoV) and will have unprecedented surveying power. The range of key science to be tackled by the SKA covers the epoch of re-ionization, galaxy evolution, dark energy, cosmic magnetism, strong field tests of gravity, gravitational wave detection, transients, proto-planetary discs, and the search for extra-terrestrial life [1].

More specifically, the SKA is to be a radio telescope with

– the sensitivity to detect and image hydrogen in the early universe through its enormous collecting area of about 1 million square metres. This will make it about 50 times more sensitive than the Extended Very Large Array (EVLA), and able to reach an rms noise level of 10 nano-Jy in an 8 hour integration for a continuum observation,
– 50% of the collecting area concentrated in the central 5 km diameter for optimal detection of hydrogen, pulsars, and magnetic fields,
– a fast surveying capability over the whole sky through its sensitivity and very large angle FoV. The SKA will be 10000 times faster than the EVLA in surveying the sky,
– the capability for detailed imaging of compact objects like active galactic nuclei through its large physical extent of at least 3000 km,
– a frequency range from 70 MHz to 10 GHz,
– data transport to the central data processor via very wideband (terabit/sec) fibre links.

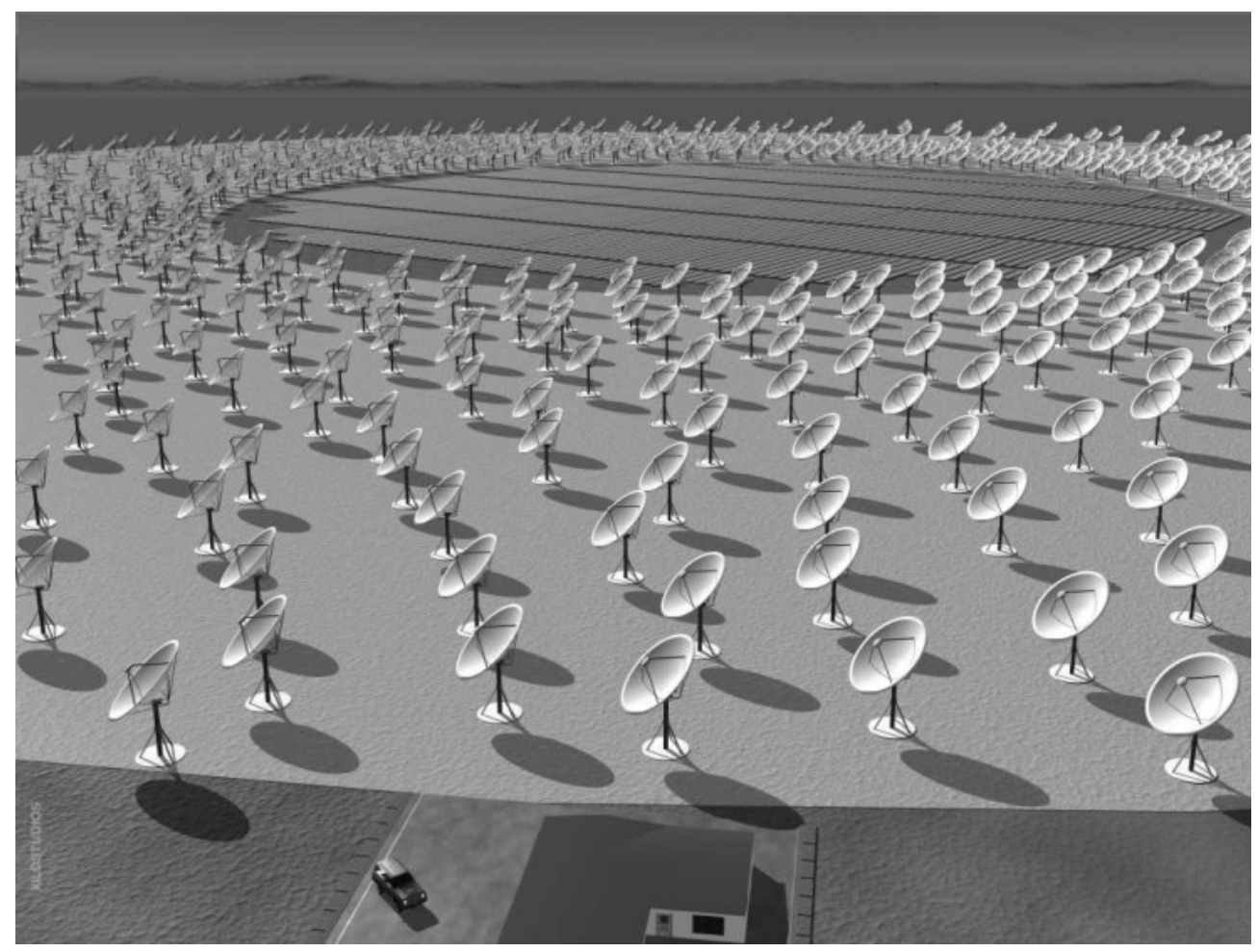

**Figure 1** – Artist's impression of the SKA central core.

The SKA is an interferometer array capable of imaging the radio sky at frequencies from 70 MHz to 10 GHz, and providing an all-sky monitoring capability at frequencies below 1.4 GHz. It covers the frequency band with three different kinds of receiving systems [2]-[4]:

1. *Sparse aperture arrays (AA-lo)* for 70-450 MHz,
2. *Dense aperture array tiles (AA-hi)* in the core of the array for all-sky monitoring in the frequency range 0.4-1.4 GHz,
3. *A 15m parabolic dish array*, with wideband single-pixel feeds for 1.2-10 GHz.

These three components all make use of the same data communications, processing, and software. An artist's impression of the array central core is shown in Figure 1.

The overall structure of the SKA signal processing system is shown in Figure 2 and includes the collector systems on the left, communications and control network in the centre and correlation and processing on the right [4].

A key cost driver for the AAs is the highest frequency supported, due to each element having an effective area which is a function of $\lambda^2$, hence, the number of elements required for a given sensitivity increases quadratically with frequency.

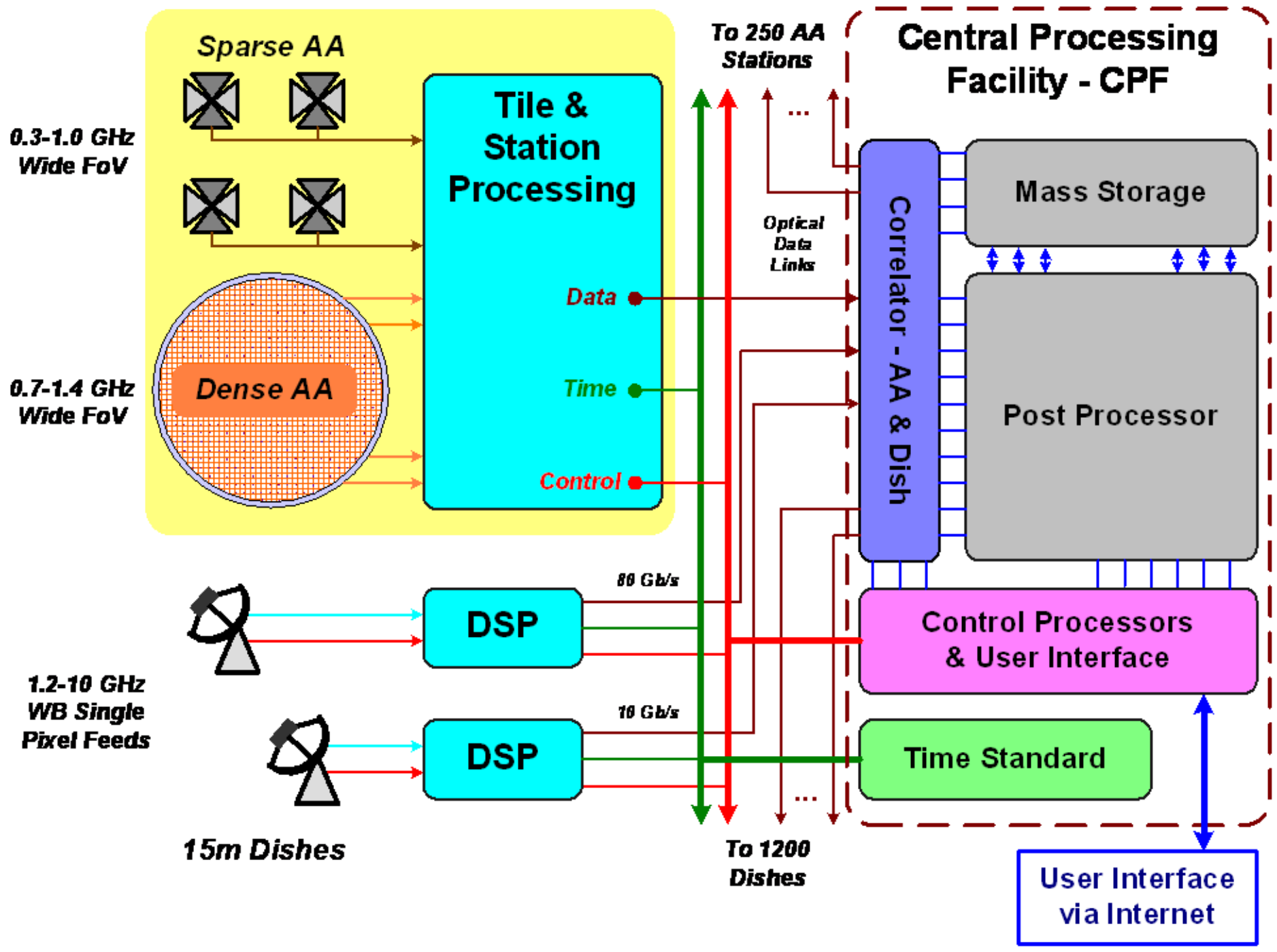

**Figure 2** – SKA implementation using AAs and single pixel feeds on dishes

Each AA-hi station is a dense ~50m diameter circular plane array consisting of $N_A \approx 75{,}000$ dual polarisation elements. An outline design of the AA system is shown in Figure 3. The design consists of four main blocks:

1. *The front-end collectors*. Each element of the AA-hi and AA-lo is positioned as part of the array design and tightly designed with its associated LNA for the lowest noise front-end design. This is amplified and passed to the "Tile processor" for initial beamforming.

2. *Tile processors* perform the first stage of beamforming, where ~64 dual polarisation elements for the AA-hi are arranged using the most effective mix of RF and digital techniques, to form a number of tile beams. The bandwidth between the Tile Processors and the Station Processors is a key determinant of the performance of the AAs.

3. *Station processors*. These bring together the output of all the AA tiles. They form the beams for transmission to the correlator. The calibration algorithms to form high precision station beams will be handled primarily by the station processors.

4. *The control processors* keep the operation of the station coupled to the rest of the SKA. They also monitor the health of the arrays, detect non-functioning components and adjust the calibration parameters appropriately.

Again, the signal processing is implemented in a two stage structure, the Tile Processors of the first stage perform the initial digitisation and beamforming, the number of primary beams - $N_B$, their frequencies plus bandwidth and bits per sample can be configured at the observation time. The constraint is the total data rate of the internal digital links.

Identical beams from all the tiles in each array are then combined in the Station Processors to produce the required station beams. To achieve the total FoV needed there are many hundreds of individual station beams. Every first stage processor has a link to all the Station Processors. The Station Processors each link to wide area communication fibres directly to the central correlator.

The signal processing, which will be using integer (fixed-point) arithmetic for efficiency in power and silicon, requires a capability estimated for an AA-hi array of ~10 Peta-MACs. It has become clear that this performance level is achievable using either a dedicated ASIC solution, or massively multi-core parallel integer processors. This latter processing solution is very attractive for flexibility and the implementation of novel algorithms.

The main objective of the paper is the design and performance optimization of a cost-effective structure of all-digital real-time space-frequency beamforming and beamsteering for the SKA station dense AA.

As mentioned above, the main challenges of the design are associated with limited computational power of a given silicon technology.

The upper bound of the computational power assessed through the product of the data bandwidth (data rate) and dynamic range (number of bits per sample) is a key determinant of all-digital beamforming performance.

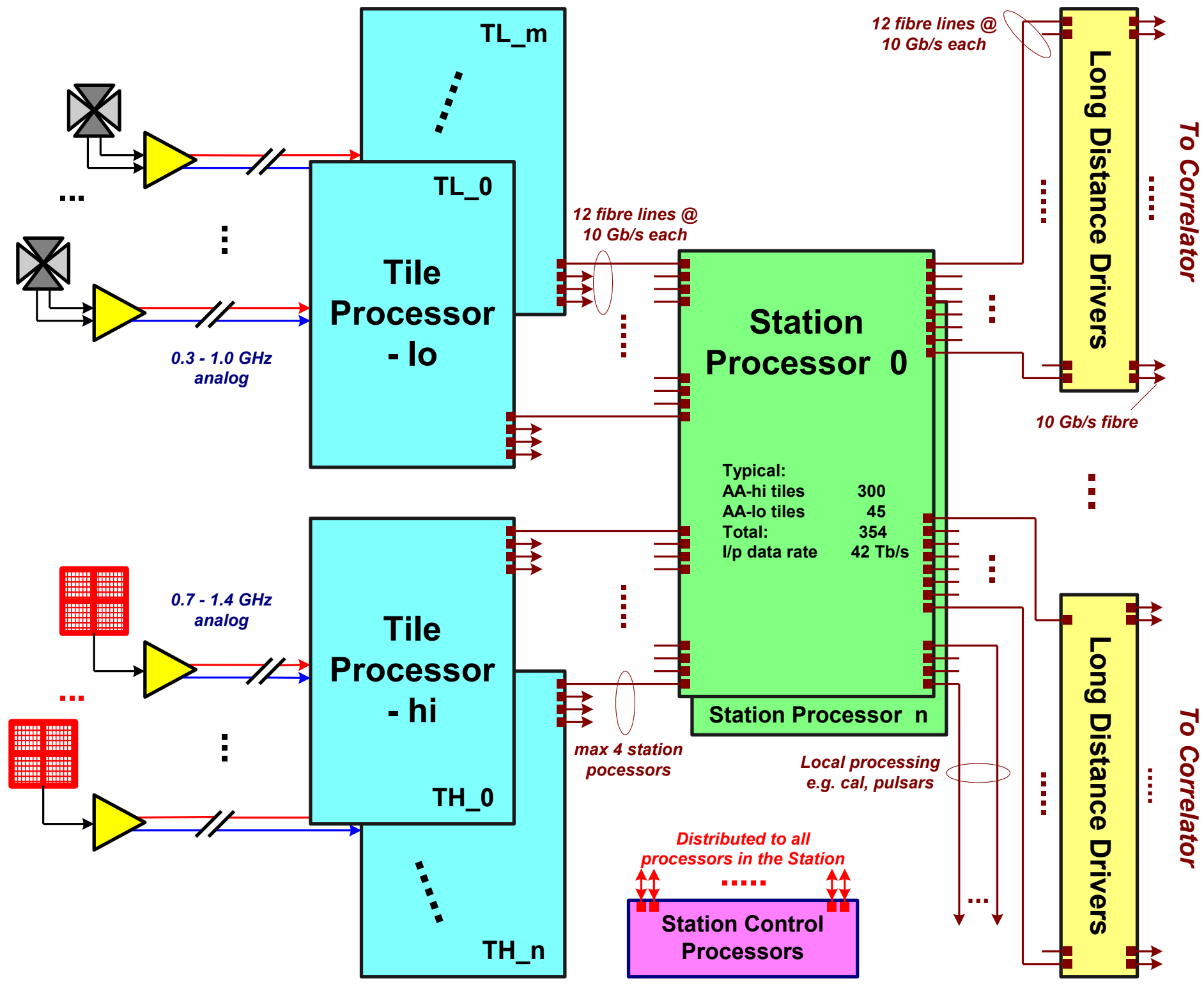

**Figure 3** – Outline AA station

According to the SKA research and design studies [2]-[4], forthcoming DSP technologies will be capable of handling the input data ranged from 4-bit at 2.4 GS/s to 6-bit at 3.4 GS/s over the 700 MHz frequency extent.

Unfortunately, these data attributes do not allow the SKA antenna processor designers to "digitize" directly the analog WB/UWB beamforming structures outlined in [5] due to enormously high computational power that would require for the all-digital implementation.

The leading idea of realizable all-digital WB beamforming underlain in the SKA station AA design [2]-[4] is to reduce the internal and output data rate by making use of parallel FFT-based frequency splitting applied directly to the AA sensor signals.

## 2. Frequency-Space Beamforming

A generalized schematic of the AA-hi beamformer that has been developed by the current stage of the SKA design [2]-[4] is shown in Figure 4 where $N_A$ is the number of antenna array sensors, $N_B$ is the number of spatial beams, $(\phi_n, \theta_n)$ is the direction the $n$-th beam ($n = 1, \ldots, N_B$) is focused on and $N_F$ is the number of frequency subbands to be specified by requirements of radio-astronomic observations.

The *Frequency-Space Beamformer* (FS-BF) depicted in the figure is a sequence of the following array signal processors: the polyphase FFT-based frequency-domain demultiplexer purposed to form $N_F$ separate frequency slots for each of $N_A$ array antenna elements, the space-domain beamformer aimed to create $N_B$ controlled spatial beams for each of $N_F$ frequency subbands, and the calibration processor to correct amplitude-phase errors in each of $N_B \times N_F$ resulting beams.

For obvious reasons, the number of linearly-independent partial beams, $N_B$ should not to be larger than the number of array antenna elements $N_A$, i.e. $N_B \le N_A$. At the current stage of the SKA design study,

$$N_A \approx 70{,}000,\ N_B \approx 1200 \text{ and } N_F = 1024 \qquad (1)$$

Let us assess the total number of fixed-point real-valued multiplications the FS-beamformer requires to process $N_F$ input successive snapshots in forming $N_B$ spatial beams for each of $N_F$ frequency subbands. This value will serve as a benchmark in estimation of a beamforming structure's implementation cost.

The number of multiplications required to perform the frequency-domain stage of the FS-beamforming is

$$M_{FS1} = O\{2 N_A N_F \log_2 N_F\} \qquad (2)$$

At the second stage of processing, when the FS-beamformer assembles and focuses the corresponding spatial beams, the number of real-valued multiplications it consumes is about

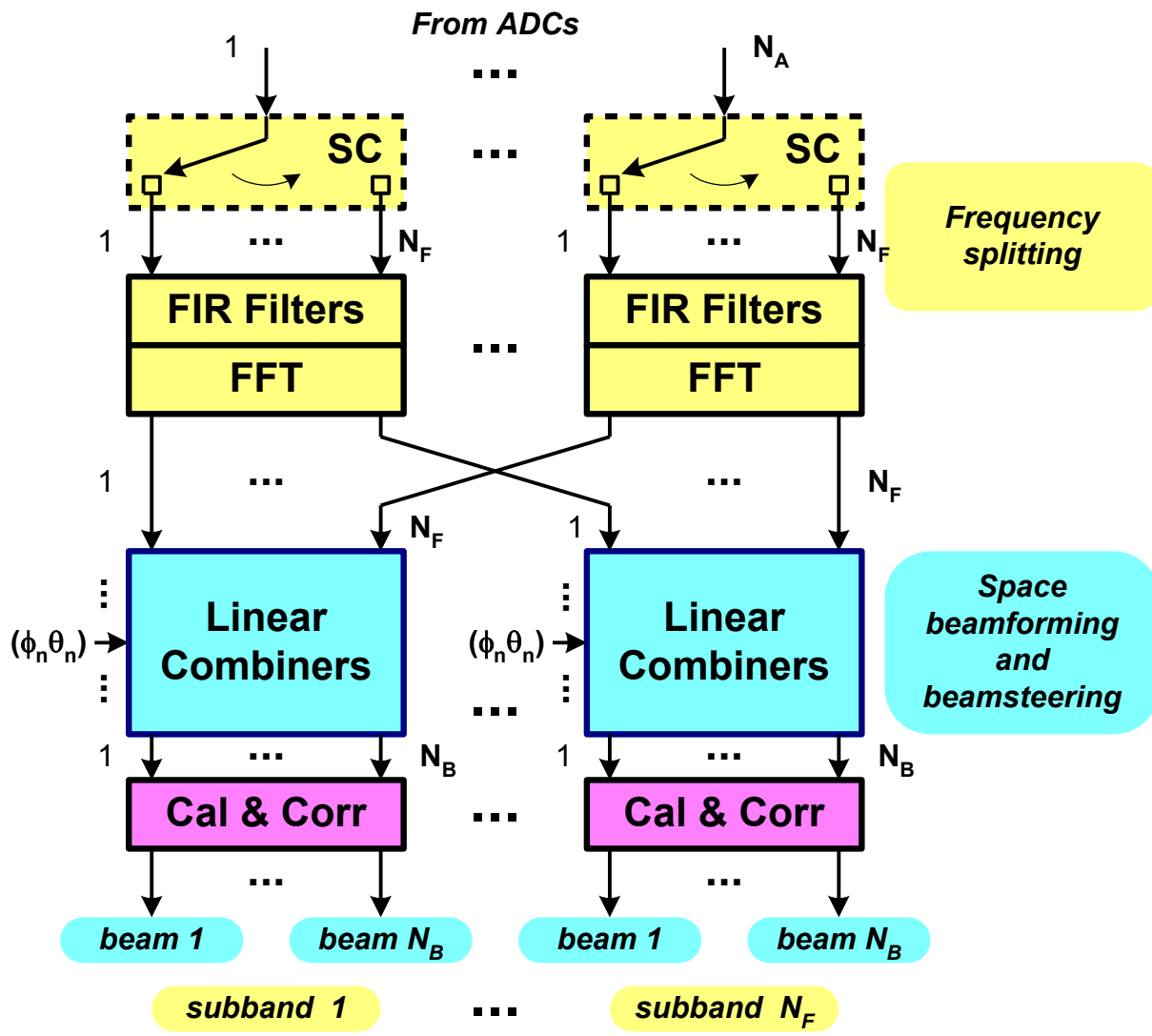

**Figure 4** – Frequency-Space beamformer

$$M_{FS2} = O\{4N_BN_AN_F\} \quad (3)$$

Thus, the total number of real-valued multiplications the FS-beamformer requires is

$$M_{FS} = M_{FS1} + M_{FS2} = O\{2N_AN_F(\log_2 N_F + 2N_B)\} \quad (4)$$

As follows from the theory of optimum space-time signal processing [6], the FS-beamforming scheme that has been derived using the "narrowband" approach [2]-[4] is a space-time filter well-matched only to the signals whose autocorrelation function widths are much longer than the maximum time delay of the signals envelopes across the array aperture.

Because of this property the beamformer may become substantially non-optimal (lossy) for a wide variety of WB/UWB cosmic signals including short-living transients, pulsars fine-grained signatures, etc.

It needs also to add the following remarks concerning HW implementations of the FS-beamformer.

Short bit-grid size quantised sample sequences are known to have the quantisation noise component attributed with the irregular multi-peak power spectrum. In order to mitigate the quantisation aliasing, we have to use an additional bank of $N_A \times N_F$ passband FIR-filters at the FFT's outputs.

The phase-shifting realization appears to be numerically expensive as it requires two operations of full dimensional complex multiplication.

At last, the FS-BF structure suffers from WB frequency aberration that significantly complicates the system of beam position control.

## 3. SPACE-FREQUENCY BEAMFORMING

Obeying the theory of optimum space-time signal processing [6], we conceive the SKA station beamformer as a wideband spatial filter to be matched to any kind of expected radio-astronomy signals through maximizing the sample signal-to-noise power ratio (SNR) at the filter outputs at each time discrete.

Thus, in order to make the space-time filter really optimum for any radio-astronomic signal, both full-band and narrowband, we have to arrange spatial beamforming and beamsteering just before frequency demultiplexing.

A newly designed structure of the all-digital wideband AA-hi beamformer that could be named as the *Space-Frequency Beamformer* (SF-BF) is presented in Figure 5.

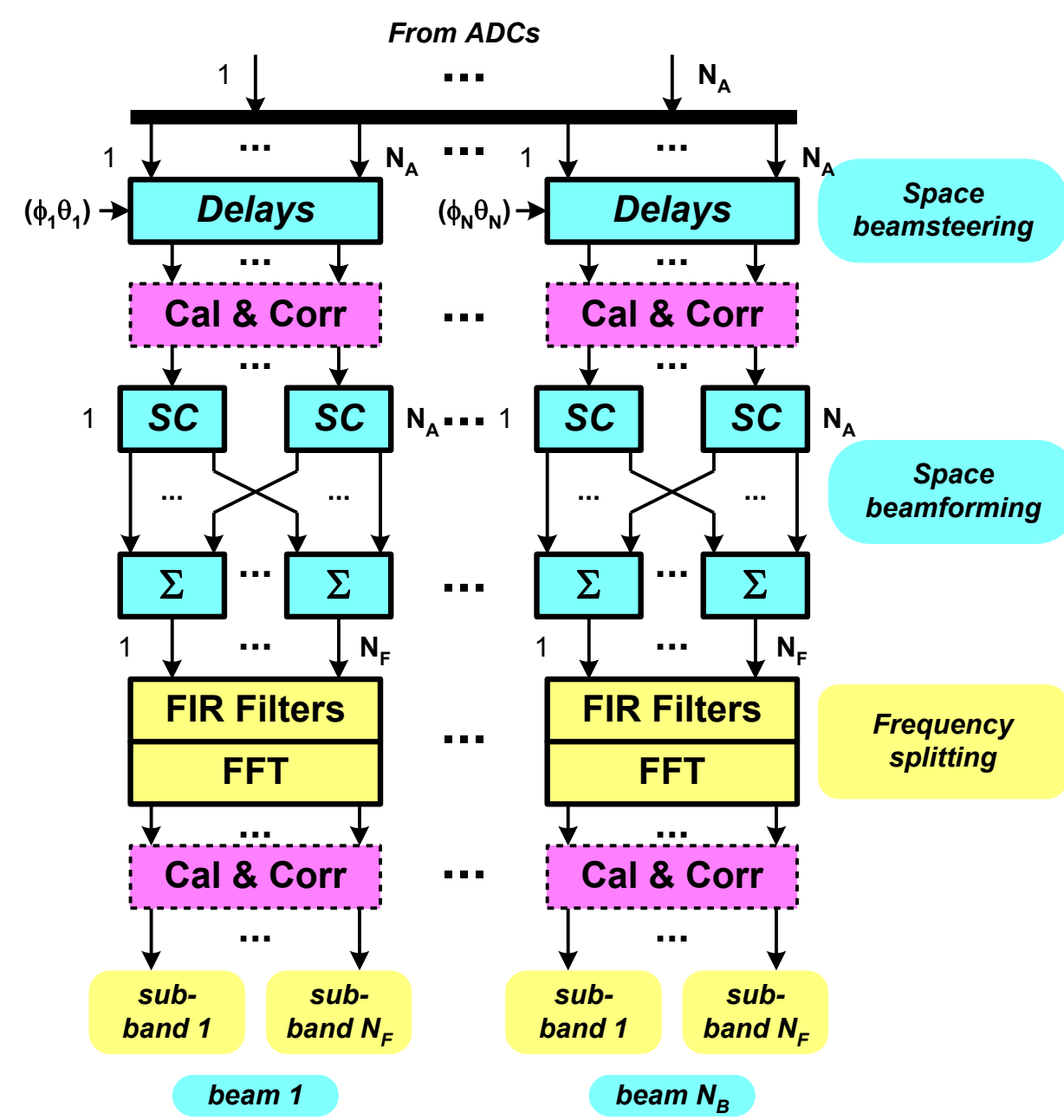

**Figure 5** – Space-Frequency beamformer

From the figure is seen that the SF-beamformer contains $N_B$ parallel processing arms to form $N_B$ individually steered wideband spatial beams, with each beam decomposed onto $N_F$ frequency subbands.

Each beamforming arm of the scheme is a two-stage real-time digital signal processor containing the wideband spatial processor to form the full-band beam pattern directed towards the prescribed point, $(\phi_n, \theta_n)$, and the polyphase FFT-based demultiplexer that performs final frequency multiple down-conversion.

Unlike the FS-BF space processing sub-system, which implements the standard narrowband routine of complex-valued matrix-matrix multiplication, the new full-band SF-

BF space processor is designed to operate in two following sub-stages:

1. *Wideband Beamsteering* realized by the $N_A$-element banks of frequency independent digital delays which align the input signal samples to focus the *n*-th SF-BF arm on the corresponding direction ($\phi_n, \theta_n$) [7],
2. *Staggered Narrowband Beamforming* implemented by the Synchronous Commutators (SC) and corresponding adders to perform time-distributed narrowband spatial integration.

Let us take a more detailed look at the both techniques of the new all-digital wideband beamforming.

**Wideband Digital Time-Aligning**

There three techniques of precise digital time-aligning could be considered eligible for use in the SKA aperture array processor:

1. *Silicon gate based keyed delay banks*,
2. *FIR-filtering* based, HW-level realizations consisted of FIFO structures for integer (coarse-grained) delaying and FIR filters for fractional (fine-grained) delaying [8],
3. *Linear interpolation*, SW-level implementations that imply re-indexing of data samples for integer delaying and time-domain linear interpolation of successive samples for fractional delaying.

The latter method looks like the most attractive to be implemented on massively multi-core parallel fixed-point (integer) processors.

The leading idea of the linear interpolation based digital delaying is very simple. Generally, a given time delay, $\tau$ is

$$\tau = \tau_I + \tau_F, \quad \tau_I = n\,\tau_S, \quad n = 1,2, \ldots \qquad (5)$$

where $\tau_S$ is the sample time discrete, $\tau_I$ and $\tau_F$ is the integer and fractional part, respectively. As mentioned above, the integer delay can easily be implemented by means of re-indexation or using a conventional shift register.

Let us consider a scenario when a reference wavefront crosses the time axis of a given space channel between two successive signal samples as shown in Figure 6.

From the picture is evident that in place of fractional delaying we can use the sample that linearly interpolated between the early and late samples using the following simple expression

$$S_T = \alpha S_L + (1 - \alpha) S_E \qquad (6)$$

where

$$\alpha = (\tau_S - \tau_I)/\tau_S = \tau_2/\tau_S \qquad (7)$$

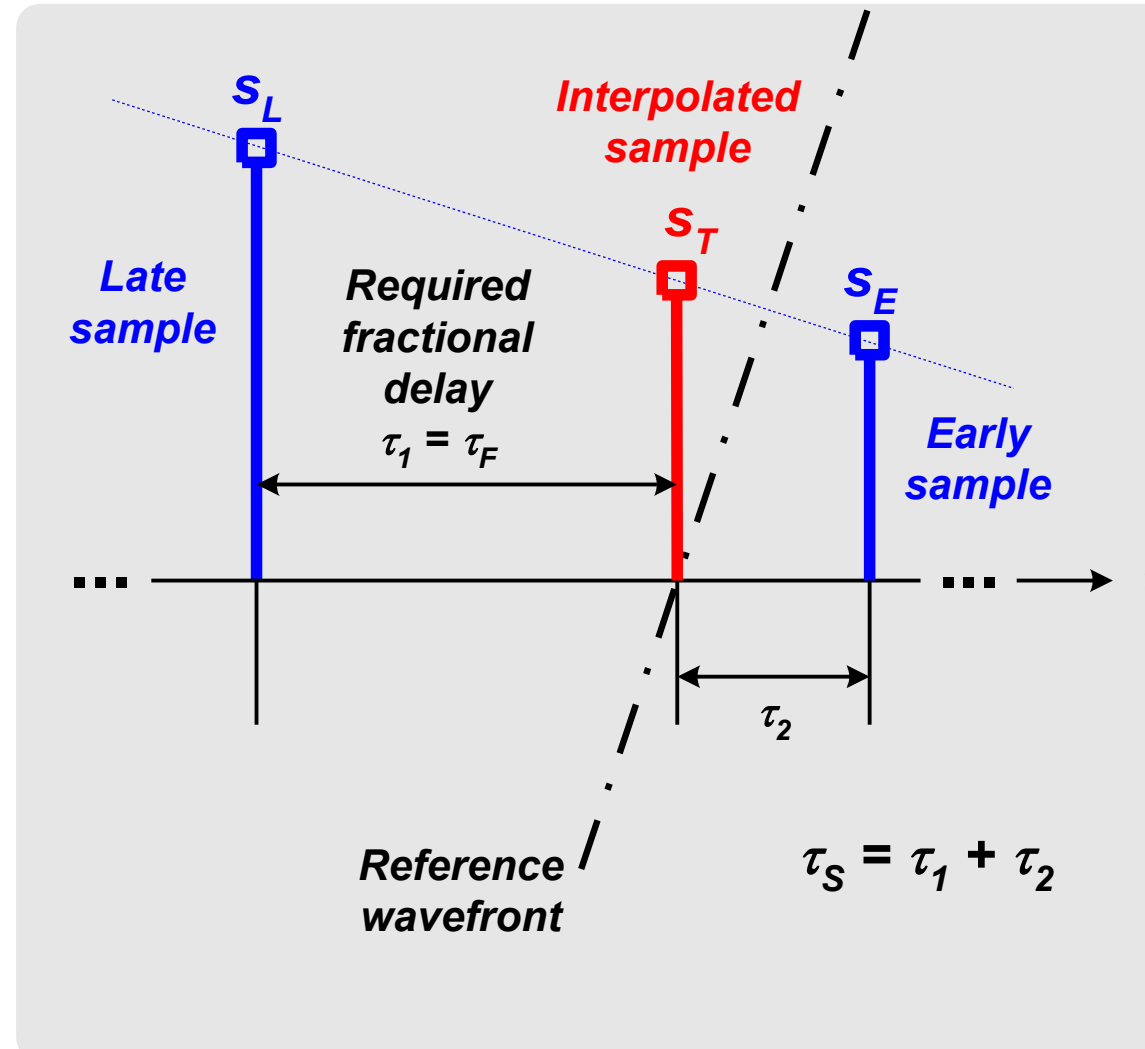

**Figure 6** – Linear interpolation based fractional delaying

Following this "substitution" technique, we can reconstruct the true wavefront position for any given boresight.

**Staggered Narrowband Beamforming**

The second stage of spatial beamforming in a single-beam SF-beamforming arm, or *beam-shaping* is implemented by $N_A$ SC's and $N_F$ adders.

The adders in the beamformer are designed to be without precision losses, i.e. the data dynamic range of partial two-item sums gradually extends from the front-end to the back-end until it becomes sufficient, e.g. 8- to 12-bit.

The digital FIR-filters and corresponding FFT units aimed to finish polyphase frequency demultiplexing in the SF-beamformer are identical to those in the FS-beamforming system.

Let us count the total number of real-valued multiplications the SF-BF consumes to form $N_B$ spatial beams for each of $N_F$ frequency slots.

If the SF-beamformer uses the linear interpolation method in time aligning the input samples then the number of real-valued multiplications at the first "space" stage of processing is about

$$M_{SF1} = O\{N_A N_B\} \qquad (8)$$

At the second "frequency" stage of processing, the FS-beamformer should perform about

$$M_{SF2} = O\{2 N_B N_F \log_2 N_F\} \qquad (9)$$

real multiplications. The total number of operations is

$$M_{SF} = M_{SF1} + M_{SF2} = O\{N_B(\mathrm{N}_A + 2N_F \log_2 N_F)\} \qquad (10)$$

## 4. Conclusions

We consider the SKA station beamformer as a wideband spatial filter to be matched to any kind of expected radio-astronomy signals through maximizing the sample SNR at the filter outputs at each time discrete.

The structure of the new beamformer is built following the rule of 'spatial integration first' to avoid processing losses in the sample SNR's for all kinds of cosmic signals in the environment of short bit-grid size digital representation of input data streams.

Beamsteering is implemented by making use of frequency independent high-precision digital time-aligning of antenna signals that reduces the complexity of the beamformer and keeps the unified time scale unbroken.

Numerous simulations have shown that as expected the SF-beamformer provides the same detection performance of weak continuous harmonic signals in additive white Gaussian noise (AWGN) as the BF-beamformer does.

Meanwhile, we expect the SF-beamformer configuration to outperform the FS-BF structure in terms of detection sensitivity for wideband and full-band signals of interest.

The newly designed beamforming configuration offers much better advantages from the standpoint of effective implementation on dedicated ASIC's, or on massively multi-core parallel integer ASIP's.

Using (1) in (4) and (10), we have the following result

TABLE I

| | **FS-beamformer** | **SF-beamformer** |
|---|---|---|
| Real-valued multiplications | **$\sim 3.6 \cdot 10^{11}$** | **$\sim 1.2 \cdot 10^{8}$** |

Providing the same signal detection capability, the new beamforming configuration outperforms by far the known structure in terms of computational complexity and hence in terms of cost, reliability and power consumption.